# Melting of superheated crystals initiates on vacancies

L. W. WANG*, Q. WANG and K. Q. LU

*Laboratory of Soft Matter Physics, Institute of Physics, Chinese Academy of Sciences,*

*Beijing 100080, China*

* Corresponding author, email: wang.lianwen@gmail.com

**Abstract**

In a large variety of ideal crystals we found that when rapidly migrating atoms squash or annihilate a neighbouring vacancy and produce a disordered cluster, the heat of migration stored in the system exceeds the enthalpy increase required for the coordinating atoms of the vacancy to form a liquid phase, i.e. *the liquid phase nucleates from vacancies*. Furthermore volumetric analysis supports this well. This vacancy-decomposition model provides quantitative information on the melting point, the latent heat and the volume change upon melting and hence clarifies the mechanism of melting.

## 1. Introduction

Since early in the last century scientists have been debating the mechanism of melting, a most common phase transition in nature [1-3]. Various models have been proposed, namely the vibrational [4], order-disorder [5], elastic [6,7], isochoric [8], defective [9,10], and entropic [11] models. However, critical issues remain unclear, i.e. *why* a crystal should melt, *how* the disordered liquid phase nucleates from the highly ordered crystalline phase, and *what* determines the latent heat and the volume change on melting? Of all the mentioned proposed models, the vibrational one seems too crude, and the others too macroscopic, to answer these questions.

Experimentally Stern [12] stressed that "because melting is a first-order phase transition, long-range fluctuations are not a significant factor and the loss of long-range order occurs by atomic mechanism which are short ranged." Recently Forsblom and Grimvall [3] demonstrated through molecular dynamics simulation in ideal crystals (crystals without surfaces and interfaces, dislocations, grain boundaries and so on, which are model crystals proposed for studying the mechanism of melting, having a theoretical melting point $T_m^K$ about 1.2 times the melting point $T_m$ of defective or real crystals, [13]) that "....the thermal fluctuation initiating melting is an aggregate typically with 6-7 interstitials and 3-4 vacancies. This mechanism differs from those that have traditionally been proposed which generally involve many more atoms at the initial melting stage". These works emphasized the decisive nature of the role that atomic fluctuations play in melting and called for new melting models having a more local perspective.

In a recent paper [14] such a microscopic model for melting of ideal crystals, namely a vacancy-decomposition model in which melting is initiated when a vacancy is squashed by neighbouring migrating atoms, has been suggested. This model answered the question *why* a crystal should melt and *what* determines the melting point.

In the present paper the vacancy-decomposition model in ideal crystals is further developed. The question of *how* the disordered liquid phase nucleates is answered and the latent heat and the volume change upon melting are quantitatively calculated

**2. Latent heat of melting**

According to [14], when an atom has energy higher than the migration energy, $E_m$, to a neighbouring vacancy, it tends to exchange position with the vacancy; see Fig. 1a. When atom A changes its position with the neighbouring vacancy, the absorbed heat of migration at the saddle point (middle) is *released* when the position-exchange process is finished (right). We call such an atom a migratable atom.

With increasing temperature, the concentration of migratable atoms increases extraordinarily, finally reaching a value of $2/N$ at $T_m^K$ (where $N$ is the number of first coordination atoms), and these will inevitably squash neighbouring vacancies (see Fig. 1b) [14]. By doing so, (1) a locally disordered atomic cluster is formed from the original highly ordered atoms *and* a vacancy and (2) the absorbed heat of migration is *stored* in the system (Fig.1 b). So it can be reasonably concluded that, for ideal crystals, the enthalpy of melting $\Delta H_m^K$ should in some way be determined by $2E_m/N$.

To be quantitative, the relationship between $\Delta H_m^K$ and $2E_m/N$ needs a detailed study. The measured $E_m$ for some bcc ($N = 8$) and some fcc/hcp ($N = 12$) structured metals were taken from [15]. As to $\Delta H_m^K$, it does not differ much from the measured entropy increase of real crystals, $\Delta H_m$, because measurements indicate that the heat capacity of a crystal and its melt become equal within $\pm 100^\circ C$ of the melting point [13], i.e. $\Delta H_m^K \approx \Delta H_m$. Values of $2E_m/N$ are plotted against $\Delta H_m$ (data taken from [16]) in Fig. 2, from which one finds that approximately:

$$\frac{2E_m}{N} = \Delta H_m \qquad (1)$$

as can be seen from the solid line in Fig. 2. Considering measurement inaccuracies in $E_m$ as well as in $\Delta H_m$, both of which are difficult to measure (in the case of W, for example, reported $\Delta H_m$ values are 52.3 kJ/mol [16] and 35.1 kJ/mol [17]), the agreement is quite good.

This agreement between $\Delta H_m$ and $2E_m/N$ indicates that in an ideal crystal, when a vacancy is squashed by two neighbouring migrating atoms, the heat of migration is sufficient for the enthalpy increase of the locally disordered atomic cluster to equal that of the liquid phase. This means that the liquid phase is nucleated among $N$ atoms near a vacancy upon squashing or annihilation of the vacancy.

## 3. Volume changes upon melting

To confirm the conclusion reached in section 2, changes in the volume occupied by the first coordination atoms to a vacancy before and after the vacancy is annihilated was calculated and compared with the measured volume changes upon melting. In the crystal state, the packing density is $\eta$ (0.68 for bcc structure and 0.74 for fcc/hcp structure), so the volume occupied by $N$ atoms is $V_N = N\Omega/\eta$ where $\Omega$ is the atomic volume. (It is noted that the volume occupied by the vacancies is neglected here in calculating the volume of the crystals because the concentration of vacancies is very low - of the order of $10^{-3}$ at the melting point [9]). When a vacancy is annihilated, its volume $V_F$ is divided among the atoms surrounding it. If, as suggested above, the liquid phase is nucleated among $N$ atoms near a vacancy, then $V_F$ is divided between those $N$ atoms. Consequently the total volume occupied by these atoms isincreased by $V_F$, i.e. $\Delta V_N = V_F$. The volume change upon melting can be calculated as

$$\frac{\Delta V_N}{V_N} = \frac{V_F}{N\Omega/\eta} \qquad (2)$$

where $N$ is the first coordination number ($N = 8$ for bcc and 12 for fcc/hcp crystals). On the basis of equation (2) and with $V_F$ values taken from [15, 18], the volume changes were calculated and compared with the measured values of $\Delta V_m/V_S$ upon melting, where $\Delta V_m$ is the volume change upon melting and $V_S$ is the volume of the crystals just before melting [17]; see Fig. 3. (Any possible differences between $\Delta V_m/V_S$ values at $T_m$ and those at $T_m^K$ are neglected here.) From Fig. 3 one finds that the volume changes calculated by equation (2) agree very well with the measured values:

$$\frac{\Delta V_N}{V_N} = \frac{\Delta V_m}{V_S} \qquad (3)$$

as shown by the solid line in Fig. 3.

To sum up, the volumetric analysis in equation (3) confirms the conclusion drawn from equation (1) that *in the melting of an ideal crystal, the liquid phase is nucleated among $N$ atoms neighbouring to a vacancy upon squashing or annihilation of the vacancy.*

### 4. Discussion

By using the vacancy-decomposition model we have been able to explain the critical issues in melting reasonably clearly and quantitatively. We have calculated the melting point [14], latent heat and the volume changes upon melting and clearly demonstrated why crystals melt, what determines the melting point and how the liquid phase develops from the highly ordered crystalline phase.

The agreement between our results and the simulation work of Forsblom *et al.* [3] is good. However, our results are more quantitative more informative.

The vacancy-decomposition model for melting connects the macroscopic properties of a crystal (melting point, latent heat and volume change) to the microscopic properties of vacancies (enthalpy of migration and volume of vacancy formation). This close connection could throw new light on research in both fields.

## 5. Conclusions

The vacancy-decomposition model for melting has been developed to a large variety of bcc, fcc/hcp structured metals. We find that when a vacancy is annihilated by neighbouring migrating atoms the heat of migration provides the enthalpy of melting, and that the volume of the vacancy corresponds to the volume change of melting, i.e. the liquid phase is nucleated among the atoms neighbouring the vacancy.

**Acknowledgements**

Financial supports from the National Natural Science Foundation of China (No. 10574145) and the National Basic Research Program of China (No. 2004CB619005) are acknowledged greatly.

**Figure captions**

**Fig. 1** Schematic diagram of atomic migration via vacancies in a crystal and corresponding heat changes in the process. (For convenience only a two-dimensional sketch is shown, although the processes we discuss in the paper are in three dimensions.)

(a) At temperatures below the melting point, atom A changes its position with the neighbouring vacancy occasionally and the absorbed heat of migration at the saddle point (middle) is *released* when the position-change process is finished (right).

(b) At the melting point, when there are always 2 out of $N$ coordination atoms that are migratable, the vacancy is squashed by the 2 migrating atoms (middle). This makes the position-change process an irreversible one (right). Consequently the heat of migration is *stored* in the system.   This is found to equal the latent heat of melting.

**Fig. 2** Comparison of the calculated energy dissipation of migrating atoms, $2E_m/N$, with the latent heat of fusion of ideal crystals, $\Delta H_m$, for bcc and fcc/hcp metals. A good coincidence between them is found.  The solid line is drawn according to equation (1).

**Fig. 3** Calculated volume changes between that occupied by $N$ coordination atoms ($V_N$) before and after a vacancy is annihilated with its volume $V_F$ ($\Delta V_N$) when divided among the $N$ atoms, and measured volume changes upon melting, $\Delta V_m/V_S$. The calculated and measured values agree very well with each other. The solid line is drawn according to equation (3).

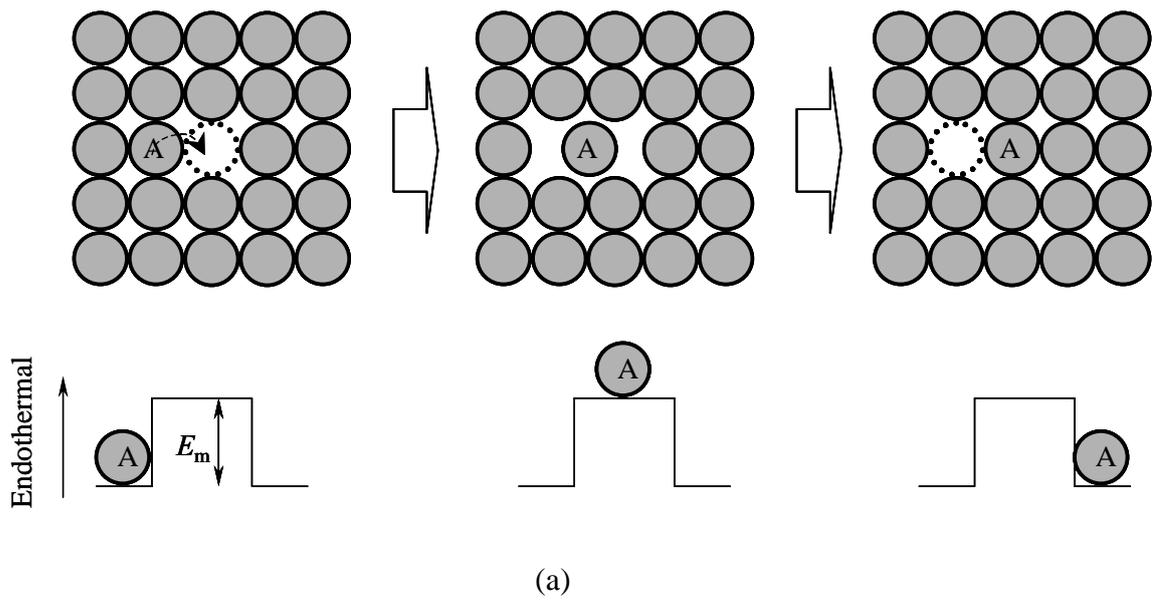

(a)

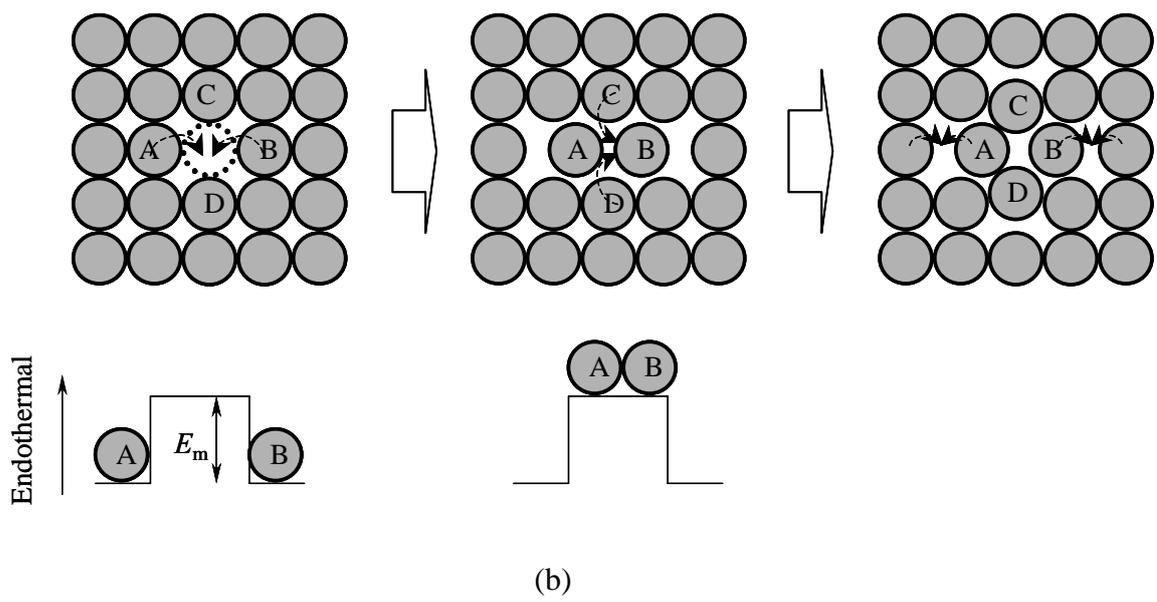

(b)

**Fig. 1/Wang et al**

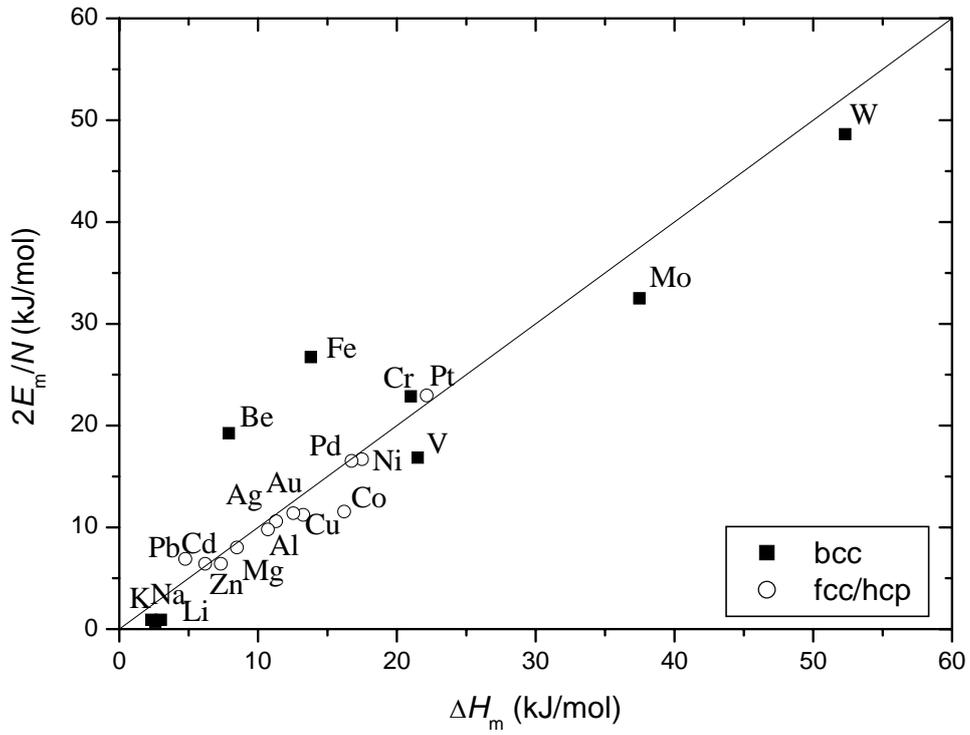

**Fig. 2/Wang et al**

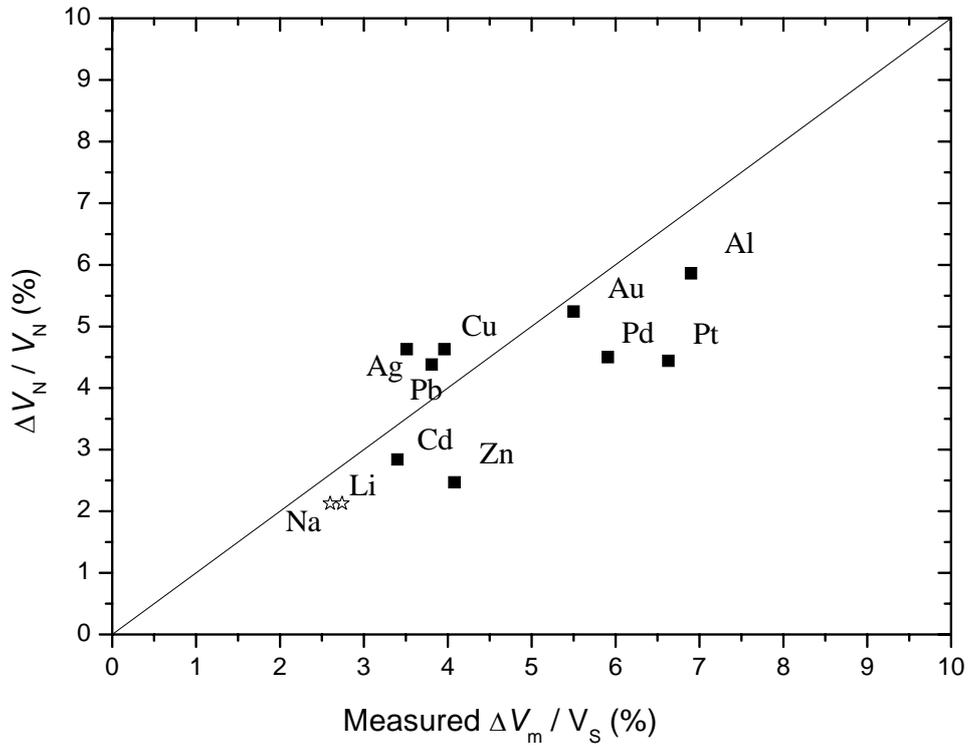

**Fig. 3/Wang et al**